\documentclass{ws-ijmpa}
\pdfoutput=1
\usepackage{physics,graphicx,mathtools}
\usepackage[super,compress]{cite}

\begin{document}

\markboth{S.R.~Ju\'{a}rez~W. and P.~Kielanowski}{Test of the 4-th
  quark generation from the CKM matrix}

\title{Test of the 4-th quark generation from the
  Cabibbo-Kobayashi-Maskawa matrix\footnote{Supported in part by
    Proyecto SIP: 20201155, Secretar\'{\i}a de Investigaci\'{o}n y
    Posgrado, Beca EDI y Comisi\'{o}n de Operaci\'{o}n y Fomento de
    Actividades Acad\'{e}micas (COFAA) del Instituto Polit\'{e}cnico
    Nacional (IPN), M\'{e}xico.}}

\author{S.R. Ju\'{a}rez Wysozka}
\address{Departamento de F\'{\i}sica, Escuela Superior de F\'{\i}sica y Matem\'{a}ticas\\
  Instituto Polit\'{e}cnico Nacional. U.P Adolfo L\'{o}pez Mateos\\
  C.P. 07738. Ciudad de M\'{e}xico, M\'{e}xico}
\author{P. Kielanowski} \address{Departamento de F\'{\i}sica, Centro
  de Investigaci\'{o}n y Estudios
  Avanzados\\
  C.P. 07000,.Ciudad de M\'{e}xico, M\'{e}xico}
% \small{\textit{E-mail:} \texttt{rebecajw@gmail.com,
% kiel@fis.cinvestav.mx}}}

\date{}

\maketitle

\begin{abstract}
  The structure of the mixing matrix, in the electroweak quark sector
  with four generations of quarks is investigated. We conclude that
  the area of the unitarity quadrangle is not a good choice as a
  possible measure of the CP~violation.  In search of new physics we
  analyze how the existence of the 4-th quark family may influence on
  the values of the Cabibbo-Kobayashi-Maskawa matrix and we show that
  one can test for the existence of the 4-th generation using the
  Jarlskog invariants of the known quarks only. The analysis based on
  the measured unitary triangle exhibits some tension with the
  assumption of three quark generations. The measurement of the
  unitarity triangle obtained from the scalar product of the second
  row/column of the CKM matrix by the complex conjugate of third
  row/column can provide information about the existence of the fourth
  generation of quarks.

  \keywords{CP-conservation; Cabibbo-Kobayashi Maskawa matrix;
    Standard Model; fourth generation of quarks.}
\end{abstract}

\ccode{PACS numbers: 12.60.-i, 11.30.Er, 12.15.Hh.}

\section{Introduction}

Some new and old experimental evidences in particle physics, represent
a challenge to be studied and explained, such as the origin of the
masses of the elementary particles and their hierarchy, the neutrino
oscillations, the asymmetry between matter and antimatter, the
presence of dark matter in the universe, etc. These issues that have
to be faced, require consideration and explanations. Enlarging the
Standard Model (SM) is one way to do it. The Standard Model (SM) of
Elementary Particles, based on the gauge invariant field theory with
the gauge group $U(1)\times SU(2)_{\text{L}}\times SU(3)_{\text{C}}$,
has had an impressive phenomenological success\cite{bib:1, bib:2,
  bib:3, bib:4, bib:5, bib:6, bib:7, bib:8, bib:9, bib:10, bib:11}
leading to the discovery of all the particles that were predicted by
the model, and describing the spectra and interactions of elementary
particles with great accuracy. Despite this success the SM has the
former drawbacks and besides has some ``anomalies'' - in the
experimental results. So, it has to be complemented with New Physics
to explain with more precision the experimental reality. For these
reasons the SM and its extensions are extensively studied in order to
remove inconsistencies and to eliminate deficiencies in its
description of elementary particles.

Motivated by the hypothesis, that is related with the fourth neutrino,
arises the interest to study an extension of the SM, by increasing the
number of families or generations and consider a fourth generation of
quarks and leptons instead of the three of the SM, and investigate in
this context, the new characteristics and consequences associated with
the CP~symmetry which is related with the asymmetry between matter and
antimatter in the universe. For a recent review of the structure of
the quark and lepton sector in the extended SM see
Ref.~\refcite{bib:34} and a very complete set of references therein.

The problem that is related with the neutrino oscillations between the
different ``flavors'' of them, that was observed by MiniBooNE in
Fermilab in 2018, encouraged the idea of extending the number of
families or flavors of the neutrinos. To explain the oscillations, the
idea (which is not discarded conclusively)\cite{bib:12, bib:12a,
  bib:12b, bib:12c} of a new neutrino, a ``sterile'' one, has been
proposed, which is a possible dark matter candidate, that besides
could be part of a fourth family of basic entities that compose the
matter and antimatter. It has been also proposed that the heavy quarks
of the fourth generation are contained in the dark matter as has been
also explored\cite{bib:13, bib:14, bib:15, bib:16, bib:17}.

The extension of the Standard Model by adding an additional family has
been studied before\cite{bib:18, bib:19} (for a broad review, see
Ref.~\refcite{bib:20} and a more recent update\cite{bib:21}). Various
phenomenological analysis examined the possibility of existence of the
fourth quark family and its compatibility with experimental
data\cite{bib:22, bib:23}.  A very detailed study of the
phenomenological consequences of the existence of the fourth
generation is contained in Ref.~\refcite{bib:24}. We use information
from that paper to study a possibility of the detection of the fourth
generation from the elements of the CKM matrix of the known quarks.

The structure of our paper is the following. The first two sections of
the paper contain introductory material and define the notation, the
next section is devoted to the description of the CKM matrix for the
extended model. Next, we discuss the unitarity of the CKM matrix and
the unitarity quadrangle. Finally we obtain the predictions for the
matrix of the Jarlskog invariants for the known quarks in the extended
model and find out how it deviates from the Standard Model. The paper
is concluded by summarizing our results and the Appendices, where we
include some details of the calculations.

\section{The Mixing Matrix in the Standard Model}

As we know, the SM has been notably successful (though the values of
the mixing parameters are not predicted by the model, neither the
masses of the elementary quarks), on the basis that there are no
experiments that are incompatible with the model. The generation of
masses in the fermionic sector of the SM (quarks and leptons) arises
through the Yukawa interactions with the Higgs doublet, using the
mechanism of spontaneous symmetry breaking.  The masses, are obtained
then, by a diagonalization of the $3\times3$ matrix in the
SM-Lagrangian's Yukawa term, and are related with the experimental
results for the quark masses.

This diagonalization, due to the unitary transformation of the
original hypothetical massless quark fields into the massive ones, has
important consequences which reverberate, in the charged interactions
of them, by the appearance of the Cabibbo-Kobayashi-Maskawa Mixing
Matrix (CKM).

A good understanding of this sector is crucial for the progress in the
theory of elementary particles and its extensions.

In the SM the Yukawa interactions are described by the three complex
$3\times 3$ matrices for the up and down quarks and charged
leptons. The charged lepton matrix is diagonal with matrix elements
proportional to the lepton masses. The matrices of the quark Yukawa
couplings are not diagonal and out of 18~complex matrix elements only
10~parameters have phenomenological significance: 6~quark masses and
4~parameters of the Cabibbo-Kobayashi-Maskawa matrix\cite{bib:25},
three angles and a phase which is non vanishing and implies that the
CP~symmetry is broken. One of the reasons for such a reduction of the
number of parameters is the freedom of choice of the phases of the
quark fields (rephasing freedom). This is the reason that all the
observables related with the Cabibbo-Kobayashi-Maskawa matrix must be
rephasing invariant.

The SM is defined by its Lagrangian. The part of the Lagrangian that
corresponds to the quark Yukawa interactions has the following form
\begin{equation}
  y_{u}\overline{u}_{R}(\phi ^{+}u_{L})+y_{d}\overline{d}_{R}(\phi d_{L})+
  \text{h.c.}
\end{equation}
Here $y_{u}$ and $y_{d}$ are the Yukawa couplings for the up and down
quarks, $\phi$ is the Higgs field and $u_{L,R}$, $d_{L,R}$ are the
quark fields. The $y_{u}$ and $y_{d}$ are $3\times 3$ complex matrices
and they are not observable in the SM. The $y_{u}$ and $y_{d}$ can be
diagonalized by biunitary transformations ($Y_{i}^{u,d}$ being their
eigenvalues)
\begin{equation}\label{eq:2}
  \operatorname{diag}(Y_{1}^{u},Y_{2}^{u},Y_{3}^{u})=U_{R}^{u}y_{u}{U_{L}^{u}}
  ^{\dagger },\quad \operatorname{diag}(Y_{1}^{d},Y_{2}^{d},Y_{3}^{d})=U_{R}^{d}y_{d}{
    U_{L}^{d}}^{\dagger }
\end{equation}
and the quark fields are transformed by the unitary transformations
$U_{L,R}^{u,d}$. Upon this transformation and after the spontaneous
symmetry breaking the terms of the Yukawa Lagrangian~ are transformed
into the quark mass terms with the quark masses equal to
$m_{i}^{u,d}=Y_{i}^{u,d}v/\sqrt{2}$ ($v$ is the Higgs field vacuum
expectation value) and the charged current ceases to be diagonal and
instead it is described by the Cabibbo-Kobayashi-Maskawa
(CKM)~$V_{\text{CKM}}$
\begin{equation}
  \begin{gathered}
    \mathcal{L}_{qW}=-\frac{g}{\sqrt{2}}\left(
      \overline{u}_{L},\overline{c}_{L}, \overline{t}_{L}\right)
    \gamma ^{\mu }W_{\mu }^{\dag }\,V_{\text{CKM}}
    \begin{pmatrix}
      d_{L} \\
      s_{L} \\
      b_{L}
    \end{pmatrix}
    +\text{h.c.}\\[4pt]
    V_{\text{CKM}}=U_{L}^{u}\left( U_{L}^{d}\right) ^{\dag }=
    \begin{pmatrix}
      V_{ud} & V_{us} & V_{ub} \\
      V_{cd} & V_{cs} & V_{cb} \\
      V_{td} & V_{ts} & V_{tb}
    \end{pmatrix}.
  \end{gathered}
\end{equation}
The matrices $U_{L,R}^{u,d}$ in Eq.~\eqref{eq:2} are not equal for the
up and down quarks and there is a rephasing freedom for the quark
fields, which is the reason of the rephasing freedom of the CKM
matrix. The quark masses (or~$Y_{i}^{u,d}$) and the elements of the
CKM matrix are measured observables in the~SM.

The elements $V_{ij}$ of the $V_{\text{CKM}}$ matrix do represent the
probability of transformation between quarks or interactions between
them, which is allowed by the weak interaction.

The most frequently used representations of the mixing matrix are the
following:
\begin{enumerate}
\item In terms of the angles
  $\theta _{12}, \theta _{13}, \theta _{23}$, and the phase $\delta$\,
  the matrix$\ V_{\text{CKM}}$ can be represented as the product of
  the following three rotations\cite{bib:1},\break where
  $c_{ij}\equiv \cos \theta _{ij}$, $s_{ij}\equiv \sin \theta _{ij}$
  are tagged by $i,j=1,2,3$; and $\delta$ is the phase.
  \begin{equation}\begin{gathered}
      R_{23}=
      \begin{pmatrix}
        1 & 0 & 0 \\
        0 & c_{23} & s_{23} \\
        0 & -s_{23} & c_{23}
      \end{pmatrix}
      ,\ \ \ R_{13}=
      \begin{pmatrix}
        c_{13} & 0 & s_{13}e^{-i\delta } \\
        0 & 1 & 0 \\
        -s_{13}e^{i\delta } & 0 & c_{13}
      \end{pmatrix}
      ,\\[12pt]
      R_{12}=
      \begin{pmatrix}
        c_{12} & s_{12} & 0 \\
        -s_{12} & c_{12} & 0 \\
        0 & 0 & 1
      \end{pmatrix}
      ,
    \end{gathered}
  \end{equation}
  as
  \begin{multline}
    V_{\text{CKM}}=R_{23}R_{13}R_{12}\\=
    \begin{pmatrix}
      c_{12}c_{13} & s_{12}c_{13} & s_{13}e^{-i\delta } \\
      -s_{12}c_{23}-c_{12}s_{23}s_{13}e^{i\delta } &
      c_{12}c_{23}-s_{12}s_{23}s_{13}e^{i\delta } & s_{23}c_{13} \\
      s_{12}s_{23}-c_{12}c_{23}s_{13}e^{i\delta } &
      -c_{12}s_{23}-s_{12}c_{23}s_{13}e^{i\delta } & c_{23}c_{13}
    \end{pmatrix}.
  \end{multline}

\item The Wolfenstein parametrization\cite{bib:26} with the
  corrections by Buras, Lautenbacher and Ostermaier\cite{bib:27} and
  the CKMfitter Group\cite{bib:27a} in terms of the parameters
  $A,\ \lambda,\ \rho$ and $\eta$, where
  \begin{equation*}
    s_{12}\equiv\lambda,\quad s_{23}\equiv A\lambda ^{2},\quad
    s_{13}e^{-i\delta }\equiv A\lambda ^{3}\left( \rho-i\eta \right),
  \end{equation*}
  is
  \begin{equation*}
    V_{\text{CKM}}\approx
    \begin{pmatrix}
      1-\frac{\lambda ^{2}}{2} & \lambda
      & A\lambda ^{3}\left(\rho-i\eta\right) \\
      -\lambda & \phantom{aa}1-\frac{\lambda ^{2}}{2}\phantom{aa} & A\lambda ^{2} \\
      A\lambda ^{3}\left( 1-\rho-i\eta\right) & -A\lambda ^{2} & 1
    \end{pmatrix}.
  \end{equation*}
  The $\lambda \approx 0.23$ is related to the Cabibbo
  angle~$\theta_{12}$ that plays an important role in the expansion
  and $\eta$ represents the imaginary part that is responsible for the
  CP~violation. In this approximation terms of order
  $\order{\lambda ^{4}}$ have been neglected .
\end{enumerate}

\section{Extended Model}

The subject of our analysis is the quark mixing matrix in the four
families extension of the SM. The forth quark family consists of the
quark doublet $(B,T)$ and the Mixing Matrix for four generations
$V_{\text{4Fam}}$ is a unitary $4\times 4$ matrix. After taking into
account the rephasing freedom it depends on 3 phases
$\delta=\delta_{13}$, $\delta _{14}$, $\delta _{24}$ and 6 rotation
angles $\theta _{12}$, $\theta _{13}$, $\theta _{14}$, $\theta _{23}$,
$\theta _{24}$, $\theta _{34}$. It can be represented by the product
of the 6 rotation-like transformations $R_{12}$, $R_{13}$, $R_{23}$,
$R_{14}$, $R_{24}$, $R_{34}$ (see Ref.~\refcite{bib:28})
\begin{gather*}
  R_{12}=
  \begin{pmatrix}
    c_{12} & s_{12} & 0 & 0 \\
    -s_{12} & c_{12} & 0 & 0 \\
    0 & 0 & 1 & 0 \\
    0 & 0 & 0 & 1
  \end{pmatrix}
  ,\ \ \ R_{13}=
  \begin{pmatrix}
    c_{13} & 0 & s_{13}e^{-i\delta } & 0 \\
    0 & 1 & 0 & 0 \\
    -s_{13}e^{i\delta } & 0 & c_{13} & 0 \\
    0 & 0 & 0 & 1
  \end{pmatrix},\\[6pt]
  R_{23}=
  \begin{pmatrix}
    1 & 0 & 0 & 0 \\
    0 & c_{23} & s_{23} & 0 \\
    0 & -s_{23} & c_{23} & 0 \\
    0 & 0 & 0 & 1
  \end{pmatrix},\ \ \ R_{14}=
  \begin{pmatrix}
    c_{14} & 0 & 0 & s_{14}e^{-i\delta _{14}} \\
    0 & 1 & 0 & 0 \\
    0 & 0 & 1 & 0 \\
    -s_{14}e^{i\delta _{14}} & 0 & 0 & c_{14}
  \end{pmatrix},\\[6pt]
  R_{24}=
  \begin{pmatrix}
    1 & 0 & 0 & 0 \\
    0 & c_{24} & 0 & s_{24}e^{-i\delta _{24}} \\
    0 & 0 & 1 & 0 \\
    0 & -s_{24}e^{i\delta _{24}} & 0 & c_{24}
  \end{pmatrix},\ \ \ R_{34}=
  \begin{pmatrix}
    1 & 0 & 0 & 0 \\
    0 & 1 & 0 & 0 \\
    0 & 0 & c_{34} & s_{34} \\
    0 & 0 & -s_{34} & c_{34}
  \end{pmatrix}.
\end{gather*}
in the following way:
\begin{equation}
  V_{\text{4Fam}}=R_{34}R_{24}R_{14}R_{23}R_{13}R_{12}= \begin{pmatrix}
    V_{ud} & V_{us} & V_{ub} & V_{uB}\\
    V_{cd} & V_{cs} & V_{cb} & V_{cB}\\
    V_{td} & V_{ts} & V_{tb} & V_{uB}\\
    V_{Td} &V_{Ts} &V_{Tb} &V_{TB}
  \end{pmatrix}.
\end{equation}
Here the product of $R_{23}R_{13}R_{12}$ that contains only one phase
is equivalent to the one obtained for three families of the SM. In
order to simplify the further calculations, it is described in the
following way
\begin{equation}
  V_{\text{CKM3}}=R_{23}R_{13}R_{12}=\left( 
    \begin{array}{cccc}
      M_{11} & M_{12} & M_{13} & 0 \\ 
      M_{21} & M_{22} & M_{23} & 0 \\ 
      M_{31} & M_{32} & M_{33} & 0 \\ 
      0 & 0 & 0 & 1
    \end{array}
  \right),
\end{equation}
where,
\begin{equation*}\begin{aligned}
    M_{11} &=c_{12}c_{13},\ \ \ \ M_{12}=s_{12}c_{13},\ \ \ \ \
    M_{13}=s_{13}e^{-i\delta },\   \\
    M_{21} &=-c_{23}s_{12}-s_{23}s_{13}e^{i\delta }c_{12},\ \ \
    M_{22}=c_{23}c_{12}-s_{23}s_{13}e^{i\delta }s_{12},\\
    M_{23}&=s_{23}c_{13}, \ \ \
    M_{31} =s_{23}s_{12}-s_{13}e^{i\delta }c_{12}c_{23},\\
    M_{32}&=-s_{23}c_{12}-c_{23}s_{13}e^{i\delta }s_{12},\ \ \
    M_{33}=c_{23}c_{13}.
  \end{aligned}
\end{equation*}
In the case of $R_{34}R_{24}R_{14}$ we obtain
\begin{equation}
  R_{34}R_{24}R_{14}=
  \begin{pmatrix}
    c_{14} & 0 & 0 & s_{14}e^{-i\delta _{14}} \\
    -s_{14}e^{i\delta _{14}}s_{24}e^{-i\delta _{24}} & c_{24} & 0 &
    s_{24}e^{-i\delta _{24}}c_{14} \\
    -s_{14}e^{i\delta _{14}}s_{34}c_{24} & -s_{34}s_{24}e^{i\delta
      _{24}} &
    c_{34} & s_{34}c_{24}c_{14} \\
    -s_{14}e^{i\delta _{14}}c_{34}c_{24} & -c_{34}s_{24}e^{i\delta
      _{24}} & -s_{34} & c_{34}c_{24}c_{14}
  \end{pmatrix}.
\end{equation}
The following step is to evaluate the product
$V_{\text{4Fam}}=R_{34}R_{24}R_{14}V_{\text{CKM3}}$:
\begin{multline*}
  V_{\text{4Fam}}=R_{34}R_{24}R_{14}V_{\text{CKM3}},\\
  =
  \begin{pmatrix}
    c_{14} & 0 & 0 & s_{14}e^{-i\delta _{14}} \\
    -s_{14}e^{i\delta _{14}}s_{24}e^{-i\delta _{24}} & c_{24} & 0 &
    s_{24}e^{-i\delta _{24}}c_{14} \\
    -s_{14}e^{i\delta _{14}}s_{34}c_{24} & -s_{34}s_{24}e^{i\delta
      _{24}} &
    c_{34} & s_{34}c_{24}c_{14} \\
    -s_{14}e^{i\delta _{14}}c_{34}c_{24} & -c_{34}s_{24}e^{i\delta
      _{24}} & -s_{34} & c_{34}c_{24}c_{14}
  \end{pmatrix}
  \begin{pmatrix}
    M_{11} & M_{12} & M_{13} & 0 \\
    M_{21} & M_{22} & M_{23} & 0 \\
    M_{31} & M_{32} & M_{33} & 0 \\
    0 & 0 & 0 & 1
  \end{pmatrix},
\end{multline*}
which gives
\begin{multline}
  V_{\text{4Fam}} = \\
  \left(
    \begin{smallmatrix}
      c_{14}M_{11} & c_{14}M_{12} & c_{14}M_{13} & s_{14}e^{-i\delta _{14}} \\[4pt]
      \substack{
        -s_{14}e^{i\delta _{14}}s_{24}e^{-i\delta _{24}}M_{11} \\
        +c_{24}M_{21}} & \substack{
        -s_{14}e^{i\delta _{14}}s_{24}e^{-i\delta _{24}}M_{12} \\
        +c_{24}M_{22}} & \substack{
        -s_{14}e^{i\delta _{14}}s_{24}e^{-i\delta _{24}}M_{13} \\
        +c_{24}M_{23}}
      & c_{14}s_{24}e^{-i\delta _{24}} \\[4pt]
      \substack{
        -c_{24}s_{14}s_{34}e^{i\delta _{14}}M_{11} \\
        -s_{24}s_{34}e^{i\delta _{24}}M_{21} \\
        +c_{34}M_{31}} & \substack{
        -c_{24}s_{14}s_{34}e^{i\delta _{14}}M_{12} \\
        -s_{24}s_{34}e^{i\delta _{24}}M_{22} \\
        +c_{34}M_{32}} & \substack{
        -c_{24}s_{14}s_{34}e^{i\delta _{14}}M_{13} \\
        -s_{24}s_{34}e^{i\delta _{24}}M_{23} \\
        +c_{34}M_{33}}
      & c_{14}c_{24}s_{34} \\[4pt]
      \substack{
        -c_{24}s_{14}c_{34}e^{i\delta _{14}}M_{11} \\
        -s_{24}c_{34}e^{i\delta _{24}}M_{21} \\
        -s_{34}M_{31}} & \substack{
        -c_{24}s_{14}c_{34}e^{i\delta _{14}}M_{12} \\
        -s_{24}c_{34}e^{i\delta _{24}}M_{22} \\
        -s_{34}M_{32}} & \substack{
        -c_{24}s_{14}c_{34}e^{i\delta _{14}}M_{13} \\
        -s_{24}c_{34}e^{i\delta _{24}}M_{23} \\
        -s_{34}M_{33}} & c_{14}c_{24}c_{34}
    \end{smallmatrix}
  \right).
\end{multline}

\section{Explicit unitarity for n = 4}

To simplify notation, from now on, we consider that
$V=V_{\text{4Fam}}$.

The unitarity relations are described by the following 2~equations:
\begin{equation}
  \sum_{l=u,c,t,T}V_{lk}V_{lj}^{\ast
  }=\sum_{l=d,s,b,B}V_{kl}V_{jl}^{\ast }=\delta _{kj},
\end{equation}
which give 6~relations for $k\neq j$
\begin{equation}\begin{aligned}
    \sum\limits_{j=d,s,b,B}V_{uj}V_{cj}^{\ast } &=&0,\ \ \
    \sum\limits_{j=d,s,b,B}V_{uj}V_{tj}^{\ast }=0,\ \ \
    \sum\limits_{j=d,s,b,B}V_{uj}V_{Tj}^{\ast }=0,  \\
    \sum\limits_{j=d,s,b,B}V_{cj}V_{tj}^{\ast } &=&0,\ \ \
    \sum\limits_{j=d,s,b,B}V_{cj}V_{Tj}^{\ast }=0,\ \ \
    \sum\limits_{j=d,s,b,B}V_{tj}V_{Tj}^{\ast }=0.
  \end{aligned}
\end{equation}
In particular, we consider $k=u$, $j=c$
\begin{equation}
  V_{ud}^{\ast }V_{cd}+V_{us}^{\ast }V_{cs}+V_{ub}^{\ast }V_{cb}+V_{uT}^{\ast
  }V_{cT}=0.
\end{equation}
As a consequence we find that
\begin{multline}
  V_{ki}\left( \sum\limits_{l=d,s,b,B}V_{kl}V_{jl}^{\ast }\right)
  V_{ji}^{\ast }=\delta _{kj}V_{ki}V_{ji}^{\ast}\\ \Rightarrow
  \operatorname{Im}V_{ki}\bigg( \sum\limits_{l\neq
    i}V_{kl}V_{jl}^{\ast }\bigg) V_{ji}^{\ast }=0,\ \ \ j\neq k.
\end{multline}
As an example for $i=d$, $j=c$ and $k=u$ we have
\begin{gather}\label{eq:14}
  V_{ud}\left( V_{ud}^{\ast }V_{cd}+V_{us}^{\ast }V_{cs}+V_{ub}^{\ast
    }V_{cb}+V_{uT}^{\ast }V_{cT}\right) V_{cd}^{\ast }=0,
  \intertext{and} \label{eq:15}\operatorname{Im}\left(
    V_{ud}V_{cs}V_{us}^{\ast }V_{cd}^{\ast }+V_{ud}V_{cb}V_{ub}^{\ast
    }V_{cd}^{\ast }+V_{ud}V_{cT}V_{uT}^{\ast }V_{cd}^{\ast }\right)
  =0.
\end{gather}
Note that each term in Eq.~\eqref{eq:15} is of the type
$V_{\alpha j}V_{\beta k}V_{\alpha k}^{\ast }V_{\beta j}^{\ast }$ and
by construction is invariant under the rephasing of the quark fields.
Then, using the results from the Appendix~B we have obtained that the
imaginary part of the sum of three phase invariant elements are zero,
which demonstrates that the parametrization that we use is compatible
with unitarity..

\section{Unitarity quadrangle}

As an example, let us consider the representation of the following
unitarity relation\cite{bib:29, bib:30, bib:31}:
\begin{equation}\label{eq:16}
  V_{ud}V_{cd}^{\ast }+V_{us}V_{cs}^{\ast }+V_{ub}V_{cb}^{\ast
  }+V_{uB}V_{cB}^{\ast }=0.
\end{equation}
which is represented in the complex plane as a quadrangle. The
elements of the CKM matrix~$V$ in Eq.~\eqref{eq:16} are labeled by the
quark indices: $u$, $c$, $t$, $T$ for the \textit{up} quarks and $d$,
$s$, $b$, $B$ for the \textit{down} quarks.
\begin{figure}[ht]
  \centering \includegraphics[width=0.55\linewidth]{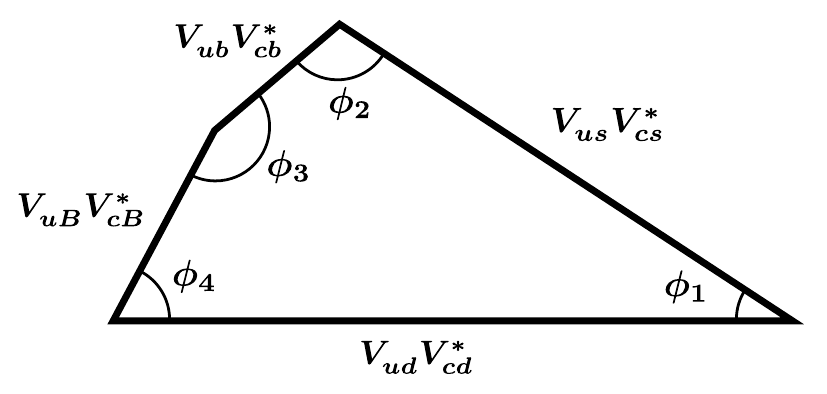}
  \caption{Unitary quadrangle.}
  \label{fig:1}
\end{figure}

\noindent The angles $\phi_{i}$ in Fig.~\ref{fig:1} are given by
\begin{equation}
  \begin{aligned}
    & \phi _{1}=\arg \left( \frac{V_{ud}V_{cd}^{\ast
        }}{V_{us}V_{cs}^{\ast }} \right) , & \phi _{2}& =\arg \left(
      \frac{V_{us}V_{cs}^{\ast }}{
        V_{ub}V_{cb}^{\ast }}\right) ,  \\
    & \phi _{3}=\arg \left( \frac{V_{ub}V_{cb}^{\ast
        }}{V_{uB}V_{cB}^{\ast }} \right) , & \phi _{4}& =\arg \left(
      \frac{V_{uB}V_{cB}^{\ast }}{ V_{ud}V_{cd}^{\ast }}\right) .
  \end{aligned}
\end{equation}
The area of the quadrangle (see Appendix~A) is evaluated in terms of
the addition of the areas of two triangles 1, and 2 in which we
decompose the quadrangle:
\begin{description}
\item[\textbf{Triangle 1}] with sides $V_{us}V_{cs}^{\ast }$ and
  $V_{ud}V_{cd}^{\ast }$ with the angle $\phi _{1}$,
\item[\textbf{Triangle 2}] with sides $V_{ub}V_{cb}^{\ast }$ and
  $V_{uB}V_{cB}^{\ast }$ with the angle $\phi _{3}$.
\end{description}
The Areas of Triangles 1 and 2 are:
\begin{align}
  \text{Area}\left( 1\right) =&\frac{1}{2}\left\vert V_{ud}V_{cd}^{\ast
                                }\right\vert \left\vert V_{us}V_{cs}^{\ast }\right\vert \sin \phi
                                _{1}=\frac{1}{2}\operatorname{Im}\left( V_{ud}V_{cs}V_{us}^{\ast
                                }V_{cd}^{\ast }\right),\\
  \text{Area}\left( 2\right) =&\frac{1}{2}\operatorname{Im}\left(
                                V_{ub}V_{uB}V_{cb}^{\ast }V_{cB}^{\ast }\right) ,
\end{align}
and
\begin{equation}
  \text{Area}_{\text{ Tot}}^{\text{ }}=\frac{1}{2}\left( \left\vert \operatorname{Im}\left(
        V_{ud}V_{cs}V_{us}^{\ast }V_{cd}^{\ast }\right) \right\vert +\left\vert 
      \operatorname{Im}\left( V_{uB}V_{cb}V_{ub}^{\ast }V_{cB}^{\ast }\right) \right\vert
  \right) .
\end{equation}

In the Appendix~A we derive
\begin{align}
  &\operatorname{Im}\left( V_{ud}V_{cs}V_{us}^{\ast }V_{cd}^{\ast
    }\right) =C_{1}\sin \delta =J_{\text{Jarlskog}},\quad
    C_{1}=c_{12}c_{13}^{2}c_{23}s_{12}s_{13}s_{23}\\
  &\operatorname{Im}V_{uB}V_{cb}V_{ub}^{\ast }V_{cB}^{\ast
    }=C_{2}\operatorname{Im}\left[ e^{-i\left( \delta _{14}-\delta
    _{24}\right) }e^{i\delta }\right] =C_{2}\sin \left(
    \delta +\delta _{24}-\delta _{14}\right)
\end{align}
where
\begin{equation*}
  C_{2}=a_{1}a_{2}a_{4}a_{5}s_{13}s_{23}c_{13},
\end{equation*}
so
\begin{equation}\label{eq:24}
  \text{Area}_{\text{ Tot}}=\frac{1}{2}\left( \left\vert C_{1}\sin \delta
    \right\vert +\left\vert C_{2}\sin \left( \delta +\delta
        _{24}-\delta _{14}\right) \right\vert \right).
\end{equation}
Eq.~\eqref{eq:24} is compatible with the unitarity triangle area in
the SM with three families, $V_{uB}=V_{cB}=0$,
\begin{equation}
  \text{Area}_{\text{ Tot}}^{\text{ triang}}=\frac{1}{2}\operatorname{Im}\left(
    V_{ud}V_{cs}V_{us}^{\ast }V_{cd}^{\ast }\right) =\frac{1}{2}J.
\end{equation}
Here $J$ is the Jarlskog invariant, which indicates, that the
CP-symmetry is broken, if it is non vanishing
\begin{equation}
  J=c_{12}c_{13}^{2}c_{23}s_{12}s_{13}s_{23}\sin \delta ,\quad J\simeq
  A^{2}\lambda ^{6}\overline{\eta }.
\end{equation}

\begin{figure}[ht]
  \centering{ \includegraphics[width=0.7\linewidth]{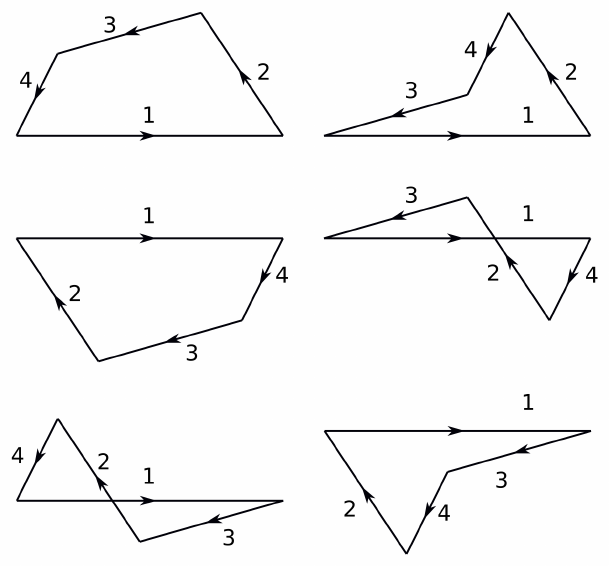}
    \caption{\label{fig:2} The different shapes of a unitarity
      quadrangle. The numbers show how the sides are ordered.}}
\end{figure}

The area of the triangle (or the Jarlskog invariant) for the case of
the three generations is the measure of the CP~violation. For the case
of four generations the situation is not so clear. If the area of the
quadrangle is not zero this implies the presence of the CP~violation
in the CKM matrix, but the area of the quadrangle cannot be
interpreted as the measure of the CP~violation. This can be seen in
Fig.~\ref{fig:2}, where it is shown that the sides of the quadrangle
can be ordered in six possible ways. Similar conclusion has been
obtained in Ref.~\refcite{bib:33} for the four neutrino mixing. This
results in three shapes of the quadrangle and the remaining three
quadrangles are the mirror images. From Fig.~\ref{fig:2} it is thus
clear that the area of a unitarity triangle cannot be uniquely defined
and thus it cannot serve as a measure of the CP~violation.

\section{CKM matrix observable effects of the fourth generation}
\label{sec:unitarity}

The CKM matrix is unitary by construction. This is true for three and
four quark generations. The existence of the fourth generation is
hypothetical and we will numerically investigate the influence of the
fourth generation on the properties of the CKM matrix for three
generations, which in such a case is a $3\times3$ submatrix of the
$V_{\text{4Fam}}$~unitary matrix. If the fourth row and column of
the~$V_{\text{4Fam}}$ are not trivial (are not equal to $(0,0,0,1)$),
then the $3\times3$ submatrix is not unitary and this has observable
consequences. We will numerically analyze the $3\times3$ submatrix
assuming that the parameters $\theta_{12}$, $\theta_{23}$,
$\theta_{13}$, $\delta$ are those obtained from the best fit for the
CKM matrix\cite{bib:1}, which are (the units of angles are radians)
\begin{equation}
  \label{eq:26}
  \theta_{12}=0.2265,\,\, \theta_{23}=0.04216,\,\, \theta_{13}=0.00364,\,\, \delta=1.2405.
\end{equation}
In our numerical analysis of the $4\times 4$ CKM~matrix we will
consider the values of the hypothetical parameters $\theta_{34}$,
$\theta_{24}$, $\theta_{14}$, $\delta_{14}$ and $\delta_{24}$ given in
Table~\ref{tab:27}. The cases A, B, C ad D were chosen in such a way
that there is a hierarchy for the mixing angles $\theta_{i4}$ similar
to that in the case of 3 generations and for the angle $\theta_{34}$
we have chosen a very conservative value~$0.01$, which means smaller
mixing that in the case of 3~generations. The values of the
CP-violating phases $\delta_{i4}$ do not have to be small and we
considered various combinations of their values in order to find out
if it would be possible to determine which pattern of the CP~violation
is preferred. For the cases E, F and~G the values of angles
$\theta_{i4}$ and phases $\delta_{i4}$ were motivated by the results
of~Ref.~\refcite{bib:24}.
\begin{table}[ph]
  \tbl{The values of the angles $\theta_{34}$, $\theta_{24}$,
    $\theta_{14}$ and the phases $\delta_{14}$, $\delta_{24}$ used for
    the numerical analysis of the observables of the CKM matrix for
    four generations.  All values of the angles and phases are given
    in radians.}
  {\begin{tabular}{llllll}
     {Case A:} &$\theta_{34}=0.01,$ &$\theta_{24}=0.001$, &$\theta_{14}=0.0001$, &$\delta_{14}=0,$ &$\delta_{24}=0$,\\
     {Case B:} &$\theta_{34}=0.01$, &$\theta_{24}=0.001$, &$\theta_{14}=0.0001$, &$\delta_{14}=1$, &$\delta_{24}=0$,\\
     {Case C:} &$\theta_{34}=0.01$, &$\theta_{24}=0.001$, &$\theta_{14}=0.0001$, &$\delta_{14}=0$, &$\delta_{24}=1$,\\
     {Case D:} &$\theta_{34}=0.01$, &$\theta_{24}=0.001$, &$\theta_{14}=0.0001$, &$\delta_{14}=1$, &$\delta_{24}=1$,\\
     {Case E:} &$\theta_{34}=0.25$, &$\theta_{24}=0.05$, &$\theta_{14}=0.02$, &$\delta_{14}=1$, &$\delta_{24}=1$,\\
     {Case F:} &$\theta_{34}=0.25$, &$\theta_{24}=0.05$, &$\theta_{14}=0.0125$, &$\delta_{14}=\frac{3}{2}\pi$, &$\delta_{24}=\frac{3}{2}\pi$,\\
     {Case G:} &$\theta_{34}=0.25$, &$\theta_{24}=0.121$, &$\theta_{34}=0.03$,
 &$\delta_{14}=1$, &$\delta_{24}=1$.
  \end{tabular}\label{tab:27}}
\end{table}

The absolute values of the $V_{\text{3Fam}}$ matrix are equal
\begin{equation}
  \label{eq:28}
  \begin{pmatrix}
    0.974461 & 0.224529 & 0.00364299  \\
    0.224379 & 0.97359 & 0.0421456  \\
    0.00896436 & 0.041342 & 0.999105
  \end{pmatrix}
\end{equation}
and let us compare this matrix with the matrices of the absolute
values of the $V_{\text{4Fam}}$ for seven cases
\begingroup\allowdisplaybreaks
\begin{subequations}\label{eq:29}
  \begin{align}
    \label{eq:29a}
    &\text{Case A:}\quad
      \begin{pmatrix}
        0.974461 & 0.224529 & 0.00364299 & 0.0001 \\
        0.224379 & 0.97359 & 0.0421456 & 0.001 \\
        0.00896509 & 0.0413499 & 0.999054 & 0.00999983 \\
        0.0000552595 & 0.000582704 & 0.0100331 & 0.999949
      \end{pmatrix}\\[10pt]
    \label{eq:29b}
    &\text{Case B:}\quad
      \begin{pmatrix}
        0.974461 & 0.224529 & 0.00364299 & 0.0001 \\
        0.224379 & 0.97359 & 0.0421456 & 0.001 \\
        0.00896581 & 0.0413498 & 0.999054 & 0.00999983 \\
        0.000100905 & 0.000572439 & 0.0100334 & 0.999949
      \end{pmatrix}\\[10pt]
    \label{eq:29c}
    &\text{Case C:}\quad
      \begin{pmatrix}
        0.974461 & 0.224529 & 0.00364299 & 0.0001 \\
        0.224379 & 0.97359 & 0.0421456 & 0.001 \\
        0.00896581 & 0.0413498 & 0.999054 & 0.00999983 \\
        0.000100905 & 0.000572439 & 0.0100334 & 0.999949
      \end{pmatrix}\\[10pt]
    \label{eq:29d}
    &\text{Case D:}\quad
      \begin{pmatrix}
        0.974461 & 0.224529 & 0.00364299 & 0.0001 \\
        0.224379 & 0.97359 & 0.0421456 & 0.001 \\
        0.00896415 & 0.0413455 & 0.999055 & 0.00999983 \\
        0.000141196 & 0.000839681 & 0.0100141 & 0.999949
      \end{pmatrix}\\[10pt]
    \label{eq:29e}
    &\text{Case E:}\quad
      \begin{pmatrix}
        0.974266 & 0.224484 & 0.00364226 & 0.0199987 \\
        0.225073 & 0.972149 & 0.0420917 & 0.0499692 \\
        0.00854277 & 0.0486111 & 0.967746 & 0.247045 \\
        0.00868513 & 0.0465875 & 0.24836 & 0.967508
      \end{pmatrix}\\[10pt]
    \label{eq:29f}
    &\text{Case F:}\quad
      \begin{pmatrix}
        0.974385 & 0.224511 & 0.00364271 & 0.0124997 \\
        0.224707 & 0.972233 & 0.0420922 & 0.0499753 \\
        0.00860203 & 0.0418036 & 0.968056 & 0.247075 \\
        0.00269613 & 0.051088 & 0.247149 & 0.967626
      \end{pmatrix}\\[10pt]
    \label{eq:29g}
    &\text{Case G:}\quad
      \begin{pmatrix}
        0.974022 & 0.224427 & 0.00364135 & 0.0299955 \\
        0.226267 & 0.965659 & 0.0418332 & 0.120651 \\
        0.00860536 & 0.0625894 & 0.96734 & 0.245485 \\
        0.00316871 & 0.11497 & 0.249982 & 0.961395
      \end{pmatrix}
  \end{align}
\end{subequations}
\endgroup From Eqs.~\eqref{eq:29} we see that significant differences
appear only for the fourth row and column (we use 8~digits after
decimal point to be able to spot the differences; the experimental
precision for the measured matrix elements may be much smaller) and
the $3\times3$ submatrices (first three rows and columns) are
compatible with Eq.~\eqref{eq:28}. One can conclude that at the level
of the absolute values of the matrix elements, the fourth generation
cannot be observed, due to the experimental precision of the
measurements of the CKM~matrix.

The analysis based on the absolute values of the CKM matrix elements
is not very sensitive to phases of these matrix elements. To analyze
the sensitivity of the CKM matrix to the values of the phases we will
consider the Jarlskog invariants, which can be introduced for mixing
in quark and lepton\cite{bib:32} sectors. For 3~families the absolute
values of all Jarlskog invariants are equal. This fact follows from
the unitarity. For 4~families the $3\times3$ CKM submatrix of
observable quarks is not unitary and thus the Jarlskog invariants are
not equal. For a $3\times3$ matrix we can construct~9 Jarlskog
invariants and they form a $3\times3$ matrix. We will consider the
$3\times3$ matrix of the absolute values of the Jarlskog
invariants~$\Delta$ with matrix elements defined by
\begin{equation}
  \label{eq:30}\Delta=
  \begin{pmatrix}
    \vert\operatorname{Im}(V_{cs}V_{tb}V_{cb}^{*}V_{ts}^{*})\vert
    &\vert\operatorname{Im}(V_{cd}V_{tb}V_{cb}^{*}V_{td}^{*})\vert
    &\vert\operatorname{Im}(V_{cd}V_{ts}V_{cs}^{*}V_{td}^{*})\vert\\[7pt]
    \vert\operatorname{Im}(V_{us}V_{tb}V_{ub}^{*}V_{ts}^{*})\vert
    &\vert\operatorname{Im}(V_{ud}V_{tb}V_{ub}^{*}V_{td}^{*})\vert
    &\vert\operatorname{Im}(V_{ud}V_{ts}V_{us}^{*}V_{td}^{*})\vert\\[7pt]
    \vert\operatorname{Im}(V_{us}V_{cb}V_{ub}^{*}V_{cs}^{*})\vert
    &\vert\operatorname{Im}(V_{ud}V_{cb}V_{ub}^{*}V_{cd}^{*})\vert
    &\vert\operatorname{Im}(V_{ud}V_{cs}V_{us}^{*}V_{cd}^{*})\vert
  \end{pmatrix}.
\end{equation}
As in the previous case we will consider seven sets of parameters
defined in Table.~\ref{tab:27}. The numerical values of the matrices
$\Delta$ for these cases are equal \begingroup\allowdisplaybreaks
\begin{subequations}\label{eq:31}
  \begin{align}
    \label{eq:31a}
    \text{Case A:}\quad
    \begin{pmatrix}
      3.25504 & 3.17461 & 3.17536 \\
      3.17538 & 3.17504 & 3.17536 \\
      3.17493 & 3.17493 & 3.17493
    \end{pmatrix}\times10^{-5}\\[10pt]
    \label{eq:31b}
    \text{Case B:}\quad
    \begin{pmatrix}
      3.25694 & 3.17539 & 3.17614 \\
      3.17537 & 3.17528 & 3.17614 \\
      3.17493 & 3.17493 & 3.17677
    \end{pmatrix}\times10^{-5}\\[10pt]
    \label{eq:31c}
    \text{Case C:}\quad
    \begin{pmatrix}
      3.25489 & 3.17282 & 3.1748 \\
      3.17482 & 3.17447 & 3.1748 \\
      3.17493 & 3.17493 & 3.17309
    \end{pmatrix}\times10^{-5}\\[10pt]
    \label{eq:31d}
    \text{Case D:}\quad
    \begin{pmatrix}
      3.25679 & 3.1736 & 3.17557 \\
      3.17481 & 3.17472 & 3.17558 \\
      3.17493 & 3.17493 & 3.17493
    \end{pmatrix}\times10^{-5}\\[10pt]
    \label{eq:31e}
    \text{Case E:}\quad
    \begin{pmatrix}
      5.27739 & 4.55467 & 6.92522 \\
      3.22541 & 2.81132 & 6.91824 \\
      3.16500 & 3.1795 & 3.16573
    \end{pmatrix}\times10^{-5}\\[10pt]
    \label{eq:31f}
    \text{Case F:}\quad
    \begin{pmatrix}
      0.57196 & 2.76940 & 0.53466 \\
      3.30709 & 2.95447 & 0.53018 \\
      3.16605 & 3.17511 & 3.16650
    \end{pmatrix}\times10^{-5}\\[10pt]
    \label{eq:31g}
    \text{Case G:}\quad
    \begin{pmatrix}
      8.26636 & 3.33730 & 9.05400 \\
      3.55075 & 2.93561 & 9.05327 \\
      3.12323 & 3.17538 & 3.12586
    \end{pmatrix}\times10^{-5}
  \end{align}
\end{subequations}
\endgroup From Eqs.~\eqref{eq:31} we see that the value of the matrix
element $\Delta_{11}$ is different than the remaining ones. This
pattern is the same in all seven cases. The experimental value of the
Jarlskog invariant $J$ quoted by the PDG\cite{bib:1} is
$J=(3.18\pm0.15)\times10^{-5}$. The value of $\Delta_{11}$ exceeds the
experimental value by more than $\sim0.07\times10^{-5}$ in all the
cases, but the case F, where it is much smaller. For the cases A, B, C
and D all remaining matrix elements of the matrix $\Delta$ are
compatible with the experimental value of~$J$. For the cases E, F and
G only the matrix elements in the third row are compatible with the
experimental value of the Jarlskog invariant. It is worth to note that
the value of $\Delta_{11}$ for the cases A, B, C and D is weakly
dependent on the values of the phases $\delta_{14}$ and
$\delta_{24}$. It means that a possible deviation of $\Delta_{11}$
from the remaining matrix elements does not give indication about the
CP~violation phases of the 4-th family.

The cases E, F and G differ significantly from those A-D. The values
of the parameters $\theta_{34}$, $\theta_{24}$ and $\theta_{14}$ for
the cases E, F and G, given in Table~\ref{tab:27}, were obtained in
Ref.~\refcite{bib:24} by analyzing the experimental data for the CKM
matrix and they are \textit{tens or hundreds times bigger than those
  guessed from the hierarchy of the CKM matrix}. Such a violation of
the hierarchy only at the level of the fourth quark family would have
a strong influence on the CKM matrix of the known quarks and there
would be a significant violation of the unitarity of the CKM matrix
for 3 generations.

\begin{figure}[ht]
  \centering \includegraphics[width=\linewidth]{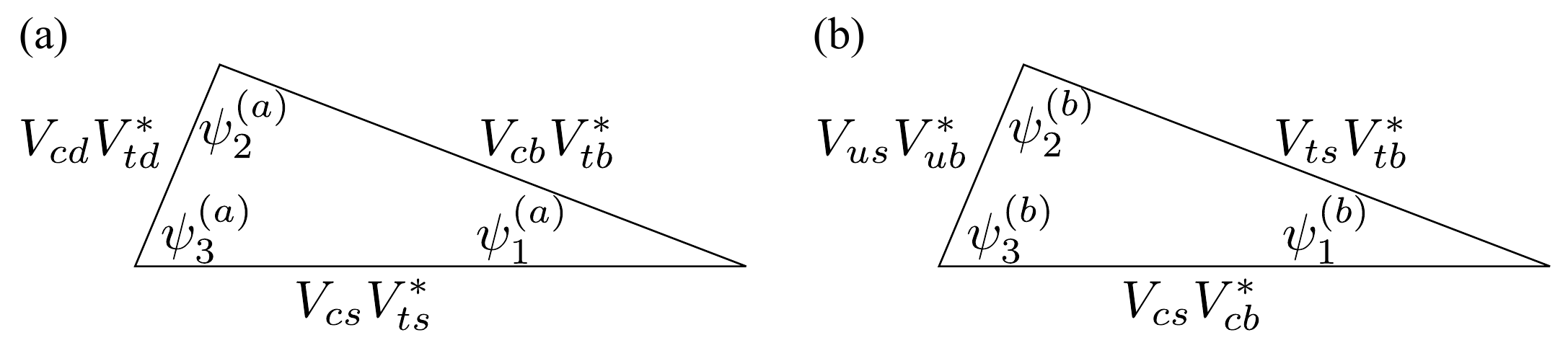}
  \caption{Unitary triangles for the determination of the phase of the
    matrix element $\Delta_{11}$. The proportions of these triangles
    do not correspond to the experimental values of the CKM matrix
    elements; the triangles with sides, proportional to the
    experimental values would be very flat and the notation in the
    drawing might be confusing. Cases (a) and (b) are explained in
    text.}
  \label{fig:3}
\end{figure}

The element $\Delta_{11}$ of the matrix $\Delta$ in Eq.~\eqref{eq:30}
can be written in the following way
\begin{equation}
  \label{eq:32}
  \Delta_{11}= \vert\operatorname{Im}(V_{cs}V_{tb}V_{cb}^{*}V_{ts}^{*})\vert = \vert(V_{cs}V_{tb}V_{cb}^{*}V_{ts}^{*})\vert\times\vert\sin(\varphi_{cs}
  +\varphi_{tb} -\varphi_{cb} -\varphi_{ts})\vert,
\end{equation}
where $\varphi_{ij}$ are the phases of the $V_{ij}$ matrix elements of
the CKM matrix. The absolute values of the matrix elements of the CKM
matrix are well measured\cite{bib:1} and for the comparison with the
experimental data one must know the relative phases of the CKM matrix
elements. The relative phases are determined from the angles of the
unitarity triangles of the CKM matrix. The measured unitarity
triangle\cite{bib:1} follows from the scalar product of the first
column of the CKM matrix by the complex conjugate of the third column
\begin{equation}
  \label{eq:33}
  V_{ud}^{\vphantom{*}}V_{ub}^{*} +V_{cd}^{\vphantom{*}}V_{cb}^{*} +V_{td}^{\vphantom{*}}V_{tb}^{*}
\end{equation}
and the angles of this triangle are equal
\begin{equation}
  \label{eq:34}
  \begin{aligned}
    \phi_{1}&=
    \arg\left(-\frac{V_{cd}^{\vphantom{*}}V_{cb}^{*}}{V_{td}^{\vphantom{*}}V_{tb}^{*}}\right)
    =\pi-\varphi_{cd}  -\varphi_{tb} +\varphi_{cb} +\varphi_{td},\\
    \phi_{2}&=
    \arg\left(-\frac{V_{td}^{\vphantom{*}}V_{tb}^{*}}{V_{ud}^{\vphantom{*}}V_{ub}^{*}}\right)
    =\pi-\varphi_{td}  -\varphi_{ub} +\varphi_{tb} +\varphi_{ud},\\
    \phi_{3}&=
    \arg\left(-\frac{V_{ud}^{\vphantom{*}}V_{ub}^{*}}{V_{cd}^{\vphantom{*}}V_{cb}^{*}}\right)
    =\pi-\varphi_{ud} -\varphi_{cb} +\varphi_{ub} +\varphi_{cd}.
  \end{aligned}
\end{equation}
From Eqs.~\eqref{eq:34} we see that the angles of the unitarity
triangle can give the information required for the determination of
the elements of the matrix $\Delta$ in Eq.~\eqref{eq:32}. The
unitarity triangles from which one can obtain the phases for the
matrix element $\Delta_{11}$ are obtained by the scalar product of the
second row of the CKM matrix by the complex conjugate of the third row
(case (a)) or by the scalar product of the second column by the
complex conjugate of the third column (case (b))
\begin{subequations}\label{eq:35}
  \begin{gather}
    \label{eq:35a}
    V_{cd}^{\vphantom{*}}V_{td}^{*} +V_{cs}^{\vphantom{*}}V_{ts}^{*} +V_{cb}^{\vphantom{*}}V_{tb}^{*}\quad\text{case (a)},\\
    \label{eq:35b}
    V_{us}^{\vphantom{*}}V_{ub}^{*} +V_{cs}^{\vphantom{*}}V_{cb}^{*}
    +V_{ts}^{\vphantom{*}}V_{tb}^{*}\quad\text{case (b)}
  \end{gather}
\end{subequations}
and are shown in Fig.~\ref{fig:3}.

\noindent The angles of these triangles are equal
\begin{subequations} \label{eq:36}
  \begin{equation}
    \label{eq:36a}
    \begin{aligned}
      \psi_{1}^{\text{(a)}}&=
      \arg\left(-\frac{V_{cs}^{\vphantom{*}}V_{ts}^{*}}{V_{cb}^{\vphantom{*}}V_{tb}^{*}}\right)
      =\pi-\varphi_{cs}  -\varphi_{tb} +\varphi_{ts} +\varphi_{cb},\\
      \psi_{2}^{\text{(a)}}&=
      \arg\left(-\frac{V_{cb}^{\vphantom{*}}V_{tb}^{*}}{V_{cd}^{\vphantom{*}}V_{td}^{*}}\right)
      =\pi-\varphi_{cb}  -\varphi_{td} +\varphi_{tb} +\varphi_{cd},\\
      \psi_{3}^{\text{(a)}}&=
      \arg\left(-\frac{V_{cd}^{\vphantom{*}}V_{td}^{*}}{V_{cs}^{\vphantom{*}}V_{ts}^{*}}\right)
      =\pi-\varphi_{cd} -\varphi_{ts} +\varphi_{td} +\varphi_{cs}.
    \end{aligned}
  \end{equation}
  \begin{equation}
    \label{eq:36b}
    \begin{aligned}
      \psi_{1}^{\text{(b)}}&=
      \arg\left(-\frac{V_{cs}^{\vphantom{*}}V_{cb}^{*}}{V_{ts}^{\vphantom{*}}V_{tb}^{*}}\right)
      =\pi-\varphi_{cs}  -\varphi_{tb} +\varphi_{cb} +\varphi_{ts},\\
      \psi_{2}^{\text{(b)}}&=
      \arg\left(-\frac{V_{ts}^{\vphantom{*}}V_{tb}^{*}}{V_{us}^{\vphantom{*}}V_{ub}^{*}}\right)
      =\pi-\varphi_{ts}  -\varphi_{ub} +\varphi_{tb} +\varphi_{us},\\
      \psi_{3}^{\text{(b)}}&=
      \arg\left(-\frac{V_{us}^{\vphantom{*}}V_{ub}^{*}}{V_{cs}^{\vphantom{*}}V_{cb}^{*}}\right)
      =\pi-\varphi_{us} -\varphi_{cb} +\varphi_{ub} +\varphi_{cs}.
    \end{aligned}
  \end{equation}
\end{subequations}
And we see that
\begin{subequations}\label{eq:37}
  \begin{align}
    \label{eq:37a}
    \Delta_{11}&= \vert\operatorname{Im}(V_{cs}V_{tb}V_{cb}^{*}V_{ts}^{*})\vert = \vert(V_{cs}V_{tb}V_{cb}^{*}V_{ts}^{*})\vert\times\vert\sin(\psi_{1}^{\text{(a)}})\vert,\\
    \intertext{and}
    \Delta_{11}&= \vert\operatorname{Im}(V_{cs}V_{tb}V_{cb}^{*}V_{ts}^{*})\vert = \vert(V_{cs}V_{tb}V_{cb}^{*}V_{ts}^{*})\vert\times\vert\sin(\psi_{1}^{\text(b)})\vert.
  \end{align}
\end{subequations}
The conclusion is that the presence of the 4-th family of quarks may
be detected by the measurement of one of the unitarity triangles in
Fig.~\ref{fig:3} and a careful comparison of the matrix elements of
the matrix $\Delta$ in Eq.~\eqref{eq:30}: if all the matrix elements
are not equal then this is a strong evidence of the existence of the
fourth quark generation.

Let us note that from the measured unitarity triangle\cite{bib:1} it
is possible to determine the three matrix elements $\Delta_{12}$,
$\Delta_{22}$ and $\Delta_{32}$ which should all be equal if there are
three generations of quarks and the $3\times3$ CKM matrix is
unitary. Their values are
\begin{equation}
  \label{eq:38}
  \begin{aligned}
    \Delta_{12}=\vert V_{cd}V_{tb}V_{cb}^{*}V_{td}^{*}\vert\times
    \vert\sin(\phi_{1})\vert&=(2.77\pm 0.18)\times 10^{-5},\\
    \Delta_{22}=\vert V_{ud}V_{tb}V_{ub}^{*}V_{td}^{*}\vert\times
    \vert\sin(\phi_{2})\vert&=(3.00\pm 0.24)\times 10^{-5},\\
    \Delta_{32}=\vert V_{ud}V_{cb}V_{ub}^{*}V_{cd}^{*}\vert\times
    \vert\sin(\phi_{3})\vert&=(3.21\pm 0.25)\times 10^{-5}
  \end{aligned}
\end{equation}
and they are compatible within an error in the value of the Jarlskog
invariant~$J$ given in Ref.~\refcite{bib:1}
\begin{equation}
  \label{eq:3}
  J=(3.00^{+0.15}_{-0.09})\times 10^{-5},
\end{equation}
but the difference between $\Delta_{12}$ and $\Delta_{32}$ exceeds one
standard deviation. The measurement of another unitary triangle may
give a more conclusive result.

\section{Results and Conclusions}

We have analyzed the structure of the CKM matrix for the extension of
the Standard Model by an additional generation of quarks. We analyzed
the properties of the unitarity quadrangles and investigated if the
area of the unitarity quadrangle might be a possible measure of the
CP~violation in the extended Standard Model. However the lack of the
uniqueness of unitarity quadrangles does not permit to use its area as
a good parameter of the CP~violation,

Assuming, that the CKM mixture of the additional quark family follows
the pattern in the $3\times3$ CKM matrix we analyze how this
additional quark family changes the mixture matrix of the known
quarks. We find that at least one of the Jarlskog invariant
($\Delta_{11}$) is significantly different than the remaining
ones. This deviation may serve as a new test of the existence of the
4-th quark family. A possibility of such a deviation is hinted by the
difference between the matrix elements $\Delta_{12}$ and $\Delta_{32}$
which exceeds one standard deviation.

Summarizing, we have analyzed what information about the fourth quark
family and new physics can be obtained from the CKM matrix of the
known quarks.

\section*{Acknowledgments}
We thank Drs. Thorsten Feldmann and Maxim Khlopov for pointing out to
us some important references.

\section*{Appendices}
\appendix
\section{} Evaluation of the Quadrangle Area.  Using the matrix
elements of the Mixing Matrix:
\begin{align*}
  & V_{ud}=a_{1}M_{11},\quad  V_{us}=a_{1}M_{12},\quad
    V_{ub}=a_{1}M_{13},\\  &V_{uB}=a_{2}e^{-i\delta _{14}},
                             V_{cd}=a_{3}e^{(i \delta_{14}-\delta_{24})}M_{11}+a_{4}M_{21},\\
  &V_{cs}=a_{3}e^{(i\delta_{14}-\delta _{24}) }M_{12}+a_{4}M_{22}, \\
  & V_{cb}=a_{3}e^{(i \delta_{14}-\delta_{24})}M_{13}
    +a_{4}M_{23},\quad  V_{cB}=a_{5}e^{-i\delta _{24}},\\
  & V_{td}=a_{6}e^{i\delta_{14}}M_{11}+a_{7}e^{i\delta_{24}}M_{21} +a_{8}M_{31},\\
  & V_{ts}=a_{6}e^{i\delta_{14}}M_{12}+a_{7}e^{i\delta _{24}}M_{22}+a_{8}M_{32}, \\
  & V_{tb}=a_{6}e^{i\delta_{14}}M_{13}+a_{7}c_{34}e^{i\delta_{24}}M_{23} +a_{8}M_{33},\quad  V_{tB}=a_{9},\\
  & V_{Td}=a_{10}e^{i\delta_{14}}M_{11}+a_{11}e^{i\delta_{24}}M_{21} +a_{12}M_{31},\\
  & V_{Ts}=a_{10}e^{i\delta_{14}}M_{12}+a_{11}e^{i\delta _{24}}M_{22} +a_{12}M_{32},\\
  & V_{Tb}=a_{10}e^{i\delta_{14}}M_{13}+a_{11}e^{i\delta_{24}}M_{23} +a_{12}M_{33},\quad V_{TB}=a_{13}.
\end{align*}
Here
\begin{align*}
  &a_{1} =c_{14}, \quad a_{2}=s_{14} \quad a_{3}=-s_{14}s_{24},
    \quad a_{4}=c_{24},\quad a_{5}=c_{14}s_{24}, \\
  &a_{6} =-s_{14}s_{34}c_{24}, \quad a_{7}=-s_{24}s_{34}, \quad a_{8}=c_{34},
    \quad a_{9}=c_{14}c_{24}s_{34}, \\
  &a_{10} =-c_{24}c_{34}s_{14}, \quad a_{11}=-c_{34}s_{24},
    \quad a_{12}=-s_{34}, \quad a_{13}=c_{14}c_{24}c_{34}.
\end{align*}
and
\begin{align*}
  &M_{11}=c_{12}c_{13},\quad M_{12}=s_{12}c_{13},\quad
    M_{13}=s_{13}e^{-i\delta },\ M_{14}=0,\\
  &M_{21}=-\left( c_{23}s_{12}+s_{23}s_{13}c_{12}e^{i\delta }\right) ,\quad
    M_{22}=c_{23}c_{12}-s_{12}s_{23}s_{13}e^{i\delta },\\
  &M_{23}=s_{23}c_{13},\ \ M_{14}=0,\quad
    M_{31}=s_{23}s_{12}-s_{13}e^{i\delta }c_{12}c_{23},\\
  &M_{32}=-s_{23}c_{12}-c_{23}s_{13}e^{i\delta }s_{12},\quad
    M_{33}=c_{23}c_{13},\quad M_{34}=0.
\end{align*}
We obtain
\begin{equation*}
  \operatorname{Im}\left( V_{ud}V_{cs}V_{us}^{\ast }V_{cd}^{\ast
    }\right) =C_{1}\sin \delta =J_{\text{Jarlskog}},\quad
  C_{1}=c_{12}c_{13}^{2}c_{23}s_{12}s_{13}s_{23}
\end{equation*}
and
\begin{align*}
  & \begin{multlined}
    V_{uB}V_{cb}V_{ub}^{\ast }V_{cB}^{\ast } =a_{1}a_{2}a_{5}e^{-i\left(
        \delta _{14}-\delta _{24}\right) }\left( a_{3}e^{i\left( \delta
          _{14}-\delta _{24}\right) }\left\vert M_{13}\right\vert
      ^{2}+a_{4}M_{13}^{\ast}M_{23}\right) \\
    =a_{1}a_{2}a_{3}a_{5}\left\vert M_{13}\right\vert
    ^{2}+a_{1}a_{2}a_{4}a_{5}s_{13}s_{23}c_{13}e^{-i\left( \delta _{14}-\delta
        _{24}-\delta\right)
    },
  \end{multlined}\\
  & \operatorname{Im}\left[ V_{uT}V_{cb}V_{ub}^{\ast }V_{cT}^{\ast
    }\right] =C_{2}e^{-i\left( \delta _{14}-\delta _{24}-\delta\right)
    },\quad
    C_{2}=a_{1}a_{2}a_{4}a_{5}s_{13}s_{23}c_{13}\\
  & \operatorname{Im}\left[ V_{uT}V_{cb}V_{ub}^{\ast }V_{cT}^{\ast
    }\right] =C_{2} \operatorname{Im}e^{-i\left( \delta _{14}-\delta
    _{24}\right) }e^{i\delta
    }=C_{2}\sin \left( \delta +\delta _{24}-\delta _{14}\right)\\
  & \text{Area}_{\text{Tot}}=\frac{1}{2}\left( \left\vert
    \operatorname{Im}\left( V_{ud}V_{cs}V_{us}^{\ast }V_{cd}^{\ast
    }\right) \right\vert +\left\vert \operatorname{Im}\left(
    V_{uB}V_{cb}V_{ub}^{\ast }V_{cB}^{\ast }\right) \right\vert
    \right)\\
  & \text{Area}_{\text{Tot}}=\frac{1}{2}\left( \left\vert C_{1}\sin
    \delta \right\vert +\left\vert C_{2}\sin \left( \delta +\delta
    _{24}-\delta _{14}\right) \right\vert \right)
\end{align*}

\section{}

The explicit evaluation of the expressions 1,2,3, from our previous
example, gives:
\begin{equation*}
  V_{ud}^{\ast }V_{cd}+V_{us}^{\ast }V_{cs}+V_{ub}^{\ast }V_{cb}+V_{uB}^{\ast
  }V_{cB}=0.
\end{equation*}
Where
\begin{equation*}
  1.-\ V_{ud}V_{cs}V_{us}^{\ast }V_{cd}^{\ast },\ \ \ 2.-\
  V_{ud}V_{cb}V_{ub}^{\ast }V_{cd}^{\ast }\text{ \ \ and \ \ }3.-\
  V_{ud}V_{cB}V_{uB}^{\ast }V_{cd}^{\ast }.
\end{equation*}
\begin{description}
\item{\textbf{Case 1.-}} $V_{ud}V_{cs}V_{us}^{\ast }V_{cd}^{\ast }$
  \begin{gather*}
    V_{ud}V_{cd}^{\ast }=a_{1}\left( a_{3}e^{-i\left( \delta
          _{14}-\delta _{24}\right) }\left\vert M_{11}\right\vert
      ^{2}+a_{4}M_{11}M_{21}^{\ast
      }\right)\\
    V_{cs}V_{us}^{\ast }=a_{1}\left( a_{3}e^{i\left( \delta
          _{14}-\delta _{24}\right) }\left\vert M_{12}\right\vert
      ^{2}+a_{4}M_{22}M_{12}^{\ast
      }\right)\\
    V_{ud}V_{cd}^{\ast }V_{cs}V_{us}^{\ast }=a_{1}^{2}\left(
      T_{1}+T_{2}+T_{3}+T_{4}\right)\\
    T_{1}=a_{3}^{2}\left\vert M_{11}\right\vert ^{2}\left\vert
      M_{12}\right\vert ^{2},\ \ \ \ \ T_{2}=a_{3}a_{4}e^{-i\left(
        \delta _{14}-\delta _{24}\right)
    }\left\vert M_{11}\right\vert ^{2}M_{22}M_{12}^{\ast },\\
    T_{3}=a_{3}a_{4}e^{i\left( \delta _{14}-\delta _{24}\right)
    }\left\vert M_{12}\right\vert ^{2}M_{11}M_{21}^{\ast },\ \ \ \
    T_{4}=a_{4}^{2}M_{11}M_{22}M_{21}^{\ast }M_{12}^{\ast }.
  \end{gather*}

  The results for the $T_{i}$ are
  \begingroup\allowdisplaybreaks\begin{gather*}
    T_{1}=\left( c_{12}c_{13}s_{12}c_{13}\right) ^{2},\ \ \ \ \operatorname{Im}T_{1}=0.\\
    T_{2}=a_{3}a_{4}e^{-i\left( \delta _{14}-\delta _{24}\right)
    }c_{12}^{2}c_{13}^{2}s_{12}c_{13}\left(
      c_{23}c_{12}-s_{12}s_{23}s_{13}e^{i\delta }\right).\\
    T_{3}=a_{3}a_{4}e^{i\left( \delta _{14}-\delta _{24}\right)
    }s_{12}^{2}c_{13}^{2}\left[ -c_{12}c_{13}\left(
        c_{23}s_{12}+s_{23}s_{13}c_{12}e^{-i\delta }\right) \right].\\
    T_{4} =a_{4}^{2}c_{12}c_{13}s_{12}c_{13}\left(
      b_{1}+b_{2}e^{i\delta
      }\right) ,\ \ \ \ \ \ b_{2}=c_{23}s_{23}s_{13}, \\
    b_{1}    =-c_{23}^{2}s_{12}c_{12}+s_{23}^{2}s_{13}^{2}s_{12}c_{12}-2c_{23}s_{23} s_{13}c_{12}^{2}\cos \delta.\\
    \intertext{and} \operatorname{Im}T_{1}=0,\\
    \operatorname{Im}T_{2}=a_{3}a_{4}c_{12}^{2}c_{13}^{3}s_{12}
    \operatorname{Im}e^{-i\left( \delta _{14}-\delta _{24}\right)
    }\left( c_{23}c_{12}-s_{12}s_{23}s_{13}e^{i\delta }\right) ,\\
    \operatorname{Im}T_{3}=-a_{3}a_{4}s_{12}^{2}c_{13}^{3}c_{12}
    \operatorname{Im}e^{i\left( \delta _{14}-\delta _{24}\right)
    }\left(c_{23}s_{12}+s_{23}s_{13}c_{12}e^{-i\delta }\right) ,\\
    \operatorname{Im}T_{4}=a_{4}^{2}c_{12}c_{13}^{2}s_{12}b_{2}e^{i\delta
    },\ \ \ \ \ \ b_{2}=c_{23}s_{23}s_{13}.
  \end{gather*}\endgroup
  Then
  \begin{multline}
    \operatorname{Im}\left[ V_{ud}V_{us}^{\ast }V_{cd}^{\ast
      }V_{cs}\right] /a_{1}^{2}
    \\
    =-c_{12}s_{12}c_{23}s_{23}s_{13}c_{13}^{2}a_{4}^{2}\operatorname{Im}e^{-i\delta
    }-c_{12}s_{12}c_{23}c_{13}^{3}a_{3}a_{4}\operatorname{Im}e^{i\left(
        \delta _{14}-\delta _{24}\right) }.
  \end{multline}

\item{\textbf{Case 2.-}} $V_{ud}V_{cb}V_{ub}^{\ast }V_{cd}^{\ast }$
  \begin{gather*}
    V_{ud}V_{cd}^{\ast }=a_{1}\left( a_{3}e^{-i\left( \delta
          _{14}-\delta _{24}\right) }\left\vert M_{11}\right\vert
      ^{2}+a_{4}M_{11}M_{21}^{\ast
      }\right)\\
    \left( V_{\text{4Fam}}\right) _{cb}=a_{3}e^{i\left( \delta
        _{14}-\delta _{24}\right) }M_{13}+a_{4}M_{23}\ ,\ \ \left(
      V_{\text{4Fam}}\right) _{ub}=a_{1}M_{13}
  \end{gather*}
  \begin{multline}
    \operatorname{Im}\left[ V_{ud}V_{cb}V_{ub}^{\ast }V_{cd}^{\ast
      }\right] /a_{1}^{2}=
    -a_{4}^{2}s_{12}s_{13}s_{23}c_{12}c_{13}^{2}c_{23}\operatorname{Im}e^{i\delta
    }\\
    -a_{3}a_{4}s_{13}c_{13}c_{12}\left[
      s_{13}c_{23}s_{12}\operatorname{Im}e^{i\left( \delta
          _{14}-\delta _{24}\right)
      }+s_{23}c_{12}\operatorname{Im}\left( e^{i\left( \delta
            _{14}-\delta _{24}\right) }e^{-i\delta }\right) \right] .
  \end{multline}

\item{\textbf{Case 3.-}} $V_{ud}V_{cB}V_{uB}^{\ast }V_{cd}^{\ast }$
  \begin{equation*}
    V_{ud}V_{cB}V_{uB}^{\ast }V_{cd}^{\ast }=a_{1}M_{11}a_{5}e^{-i\delta
      _{24}}a_{2}e^{-i\delta _{14}}\left( a_{3}e^{i\left( \delta _{14}-\delta
          _{24}\right) }M_{11}+a_{4}M_{21}\right) ^{\ast }
  \end{equation*}
  \begin{multline*}
    V_{ud}V_{cB}V_{uB}^{\ast
    }V_{cd}^{\ast }/a_{1}a_{2}a_{5}\\
    =\left( a_{3}\left( c_{12}c_{13}\right) ^{2}+a_{4}e^{i\left(
          \delta _{14}-\delta _{24}\right) }c_{12}c_{13}\left(
        -c_{23}s_{12}-s_{23}s_{13}e^{-i\delta }c_{12}\right) \right)
  \end{multline*}
  \begin{multline}
    \operatorname{Im}\left[ V_{ud}V_{cB}V_{uB}^{\ast }V_{cd}^{\ast }\right]\\
    =-a_{1}a_{2}a_{4}a_{5}\operatorname{Im}e^{i\left( \delta
        _{14}-\delta _{24}\right) }c_{12}c_{13}\left(
      c_{23}s_{12}+s_{23}s_{13}c_{12}e^{-i\delta }\right) .
  \end{multline}
\end{description}

\end{document}